\documentclass[prl,showpacs,10pt,amsmath,amssymb]{revtex4}
\usepackage{graphicx}

\begin{document}

%\draft
\title{Fractional Fokker-Planck equation for L\'evy flights in 
nonhomogeneous environments}

\author
{Tomasz Srokowski}

\affiliation{
 Institute of Nuclear Physics, Polish Academy of Sciences, PL -- 31-342
Krak\'ow,
Poland }

\date{\today}

\begin{abstract}
The fractional Fokker-Planck equation, which contains a variable 
diffusion coefficient, is discussed and solved. It corresponds to the L\'evy 
flights in a nonhomogeneous medium. For the case with the linear drift, 
the solution is stationary in the long-time limit and it represents 
the L\'evy process with a simple scaling. The solution for the drift 
term in the form $\lambda\hbox{sgn}(x)$ possesses two different scales which 
correspond to the L\'evy indexes $\mu$ and $\mu+1$ $(\mu<1)$. The former 
component of the solution prevails at large distances but it diminishes with 
time for a given $x$. The fractional moments, as a function of time, 
are calculated. They rise with time and the rate of this growth increases 
with $\lambda$.

\end{abstract}


\maketitle

Transport processes in physical systems can be very complex. The 
traditional statistical description, which relies on uniform Gaussian noises and 
position-independent transition probabilities, must fail for many realistic 
problems. A need for an alternative approach is obvious, for example, 
when one considers porous, irregular materials, containing impurities or 
entanglements which act as obstacles and dynamical traps \cite{kim}. 
Transport in a heterogeneous, in particular fractal, material involves 
position-dependent transition rates and highly complex driving forces which 
can be handled in a statistical manner (quenched disordered media) \cite{bou}. 
The Gaussian distribution not always applies. In systems which are 
characterized by long-range correlations and non-local interactions, one can 
expect the presence of long tails of the driving noise, i.e. one should 
consider the general L\'evy distributions. It is so for many physical 
phenomena \cite{shl1}, in particular 
%the distribution of distances between absorbed monomers in the polymer 
%adsorption problem \cite{gen}, as well as 
in biological \cite{wes,edw}, social \cite{bro1}, 
epidemiological problems \cite{tes}. From the Langevin equation, 
driven by the homogeneous L\'evy noise, follows the Fokker-Planck equation 
(FPE), which is fractional \cite{fog,fog1,jes,yan,den}. The importance of the general 
L\'evy distribution stems from its stability: it acts as an attractor 
in the functional space and there are no other attractors. The physical reason 
behind the L\'evy, non-Gaussian, processes traces back to the nonhomogeneous 
structure of the environment, in particular fractal or multifractal. However, 
this basis and the essential feature of the L\'evy process is only rarely 
taken explicitly into account and transport processes are described in terms 
of linear stochastic equations. The medium structure is usually reflected only 
in the form of the 
external potential or as a time-dependence of the stochastic driving. 
In this paper, we consider a nonhomogeneous L\'evy 
noise which leads to a position-dependent diffusion coefficient in FPE 
and construct the asymptotic solution of FPE.  

The problem of nonhomogeneous stochastic driving can be posed in a form 
of the Langevin equation, 
\begin{equation}
\label{lan}
dx(t)=F(x)dt+\sigma(x)dL, 
\end{equation}
with the multiplicative noise which is understood in a sense of 
the It{\^o} interpretation. Eq.(\ref{lan}) can be regarded as 
a result of the adiabatic elimination of fast variables in nonlinear processes 
with additive fluctuations. In the Gaussian case, it is suited e.g. for 
the following problems: the ensemble of two-level atoms in the electromagnetic 
field (Maxwell-Bloch equations), the parametric generation of coherent fields 
by incoming laser field, the Raman scattering, and the autocatalytic reactions 
\cite{schen}. In the present paper, we assume that the noise $L(t)$ 
is the uncorrelated L\'evy process with the stability index 
$\mu$ $(0<\mu\le2)$, the median $\gamma$, and the scale parameter $K^\mu$. 
The cumulant expansion of the characteristic function, truncated at the order 
$\mu$, produces the following fractional FPE \cite{sche}
\begin{equation}
\label{fp}
\frac{\partial}{\partial t}p(x,t)=-\frac{\partial}{\partial x}\left[(
\gamma\sigma(x)+F(x))p(x,t)\right]+K^\mu\frac{\partial^\mu}{\partial |x|^\mu}[D(x)p(x,t)] 
\end{equation}
for the probability density distribution of the variable $x$; 
in the above equation $\partial^\mu/\partial |x|^\mu$ is the Riesz fractional derivative and 
$D(x)=|\sigma(x)|^\mu$. The initial condition is $p(x,0)=\delta(x)$. 
Eq.(\ref{fp}) has been extensively studied for $D(x)=$const \cite{met}.

On the other hand, equations in the form (\ref{fp}) follow directly 
from master equation for a jumping process: 
\begin{equation}
\label{mas}
\dot p(x,t)=\int dx'[w(x|x')p(x',t)-w(x'|x)p(x,t)]. 
\end{equation}
For example, modelling the thermal 
activation of particles within the folded polymers leads to the FPE 
with the variable diffusion coefficient, which results from the polymer 
heterogeneity \cite{bro}. In general, $D(x)$ in FPE is variable if 
the transition probability $w(x|x')\ne w(|x-x'|)$. 
It is the case also for a coupled continuous time random walk (CTRW) 
which is defined in terms 
of a Poissonian waiting time distribution with a variable jumping rate 
$\nu(x)$, as well as of the L\'evy distribution of jumping size 
$Q(x)$ \cite{kam,sro}. Then $w(x|x')=\nu(x')Q(|x-x'|)$; FPE takes the form 
(\ref{fp}) with $D(x)=\nu(x)$. The parameter $\gamma$ can depend, in general, 
on the position before the jump. If it is constant, the drift term takes 
the form $\sim\gamma\nu(x)$, where 
$\gamma=\langle x\rangle_Q$. The solution of that equation for the 
power-law $\nu(x)$ and $\gamma=0$ is still the L\'evy distribution \cite{sro}. 

In this paper we demonstrate that the stability property holds also for some 
systems which contain the driving term $F(x)$. 
That is by no means obvious; for example, 
the separable solution of the fractional Schr\"odinger equation 
in Ref.\cite{bro}, which is characterized by the exponential diffusion 
coefficient and a periodic potential, looses its dependence on $\mu$ altogether, 
in the asymptotic limit. For a system with a power-law external 
potential, $|x|^c$, driven by the L\'evy noise with the constant diffusion coefficient \cite{che}, the 
stochastic properties depend on $c$: the L\'evy index in the stationary solution 
remains unchanged for the harmonic potential, whereas for larger powers 
the asymptotics is determined by $c$. As a result, the variance can be finite. 
We will solve Eq.(\ref{fp}) for the 
power-law diffusion coefficient, $D(x)=|x|^{-\theta}$ \cite{uwa1}, 
and for two simplest forms of $F(x)$ which correspond to symmetric 
potentials. 

It is convenient to handle L\'evy processes by means of
the Fox functions. If the solution of the FPE is to be the L\'evy process, 
one can expect it has the following scaling form
\begin{eqnarray} 
\label{s1}
p(x,t)=Na(t) H_{2,2}^{1,1}\left[a(t) |x|\left|\begin{array}{c}
(a_1,A_1),(a_2,A_2)\\
\\
(b_1,B_1),(b_2,B_2)
\end{array}\right.\right],
\end{eqnarray}
where $N$ is the normalization constant. The expression (\ref{s1}) is universal 
for any L\'evy process \cite{sch}. We assume the solution of Eq.(\ref{fp}) 
in this form. To find the coefficients, we require that Eq.(\ref{fp}) is 
satisfied by (\ref{s1}) in the diffusion (fluid) limit of small wave numbers $k$. 
Therefore, the solution in the form (\ref{s1}) should coincide with the exact 
solution of Eq.(\ref{fp}) for large $|x|$. 

We begin with the case $F(x)=-\lambda x$ $(\lambda>0)$ which corresponds to 
the harmonic oscillator potential. Moreover, we assume $\gamma=0$ and 
$\mu+\theta>0$. The FPE takes the form:
\begin{equation}
\label{frace}
\frac{\partial}{\partial t}p(x,t)=\lambda\frac{\partial}{\partial x}
[xp(x,t)]+K^\mu\frac{\partial^\mu}{\partial |x|^\mu}[|x|^{-\theta}p(x,t)] 
\end{equation}
and the Fourier transformation produces the result
\begin{equation}
\label{fracek}
\frac{\partial}{\partial t}{\widetilde p}(k,t)=-\lambda k\frac{\partial}{\partial k}
{\widetilde p}(k,t)-K^\mu |k|^\mu{\cal F}_c[|x|^{-\theta}p(x,t)].
\end{equation}
To solve Eq.(\ref{fracek}), we follow the procedure from 
Refs.\cite{sro,sro2}, where also the appropriate 
formulas are provided. Due to the multiplication rule, the argument of 
the Fourier transform in the last term is the Fox function of the same order 
as $p(x,t)$. We apply the formula for the cosine Fourier transform 
and expand the results in the 
fractional powers of $|k|$. In order to adjust the terms on both sides of 
Eq.(\ref{fracek}), we have to introduce conditions on the powers and to eliminate 
some of the terms by an appropriate choice of the coefficients. 
As a result, we determine the following coefficients 
of the Fox function: $a_1=1-(1-\theta)/(\mu+\theta)$, $A_1=1/(\mu+\theta)$, 
$b_2=1-(1-\theta)/(2+\theta)$, and $B_2=1/(2+\theta)$. Those values are -- 
for any choice of the other parameters -- 
a sufficient condition for Eq.(\ref{s1}) to represent the L\'evy process in 
the lowest order of the $k$-expansion: 
\begin{equation}
\label{s1ka}
{\widetilde p}(k,t)\approx 1-h_\mu a^{-\mu}|k|^\mu. 
\end{equation}
For the Fourier transform in the last term in Eq.(\ref{fracek}), we 
need to keep only the term $k^0$. We obtain 
${\widetilde p}_\theta={\cal F}_c[|x|^{-\theta}p(x,t)]\approx h_0a^\theta$, 
where $a^\theta$ results from the transformation to the scaled variable. 
The neglected terms on both sides of Eq.(\ref{fracek}) 
are of the order $|k|^{2\mu+\theta}$. 
The expansion of the Fox function around zero and $\infty$ shows that 
$(b_1,B_1)$ corresponds to the behaviour of $p(x,t)$ at $x=0$, 
whereas $(a_1,A_1)$ determine the asymptotics ($|x|\to\infty$). Therefore, 
the former ones cannot be determined in the small $k$ approximation. We assume 
values which correspond to the small $|x|$ limit of the master equation 
solution for CTRW \cite{sro2}: $b_1=\theta$ and $B_1=1$; that process is 
described by Eq.(\ref{frace}) with $\lambda=0$. 
To settle $a_2$ and $A_2$, which only weakly influence 
$p(x,t)$ in the asymptotic limit, we require that the $x$-dependence 
of the distribution (\ref{s1}) for $\lambda=0$ should coincide with 
the stretched-Gaussian exact solution in the limit $\mu\to2$ \cite{sro2}; then: 
$a_2=1/2+(1-\theta)/(2+\theta)$ and $A_2=(1+\theta)/(2+\theta)$. 
The coefficient $h_\mu$ follows 
directly from the expansion formula: $h_\mu=N(\mu+\theta)\Gamma(-\mu)
\Gamma(1+\mu+\theta)\cos(\pi\mu/2)/\Gamma[1/2+(\mu+\mu\theta+2)/(2+\theta)]
\Gamma[-(\mu+\theta)/(2+\theta)]$, 
whereas $h_0=\int_0^\infty p_\theta(ax,t)d(ax)=\lim_{s\to-1}\chi(s)=
N(\mu+\theta)/(2+\theta)\Gamma[1/2-(\theta^2+\theta-2)/(2+\theta)]$, where 
$\chi(s)$ is the Mellin transform from the Fox function. Similarly we obtain 
the normalization constant: $N=\Gamma[-\theta/(2+\theta)]\Gamma[1/2+2/
(2+\theta)]/2\Gamma(1+\theta)\Gamma[-\theta/(\mu+\theta)]$. Inserting 
${\widetilde p}$ and ${\widetilde p}_\theta$ into Eq.(\ref{fracek}) yields 
the equation for $a(t)$:
\begin{equation}
\label{eqa}
\dot a=\lambda a-K^\mu\frac{h_0}{\mu h\mu}a^{\mu+\theta+1}, 
\end{equation}
which can be solved by separation of the variables:
\begin{equation}
\label{a}
a(t)=\left[\frac{\lambda/c_L}{1-\exp[-\lambda(\mu+\theta)t]}\right]^{1/(\mu+\theta)}, 
\end{equation}
where $c_L=K^\mu h_0/\mu h_\mu$. For $\theta=0$, the above solution agrees 
with that of Ref.\cite{jes}. 

Expansion of the Fox function in powers of $1/|x|$ reveals the 
asymptotics of the L\'evy process: 
$p(x,t)\sim a(t)^{-\mu}|x|^{-\mu-1}$ for $|x|\to\infty$. $a(t)$ approaches 
with time a constant which corresponds to the stationary solution of FPE. 
The speed of that convergence depends on $\theta$: it is rapid for large $\theta$, 
whereas negative values of $\theta$ can substantially hamper the convergence. 
The meaning of the parameter $\theta$ in the context of the diffusion process 
becomes clear when we consider the case for which the variance exists, namely 
the Gaussian case $\mu=2$; it is involved 
in solution (\ref{s1}). The coefficients of the Fox function on its 
main diagonal are then equal and we can apply the reduction formula. 
The solution reads
\begin{eqnarray} 
\label{s1r}
p(x,t)=Na(t) H_{1,1}^{1,0}\left[a(t) |x|\left|\begin{array}{c}
(\frac{1}{2}+\frac{1-\theta}{2+\theta},\frac{1+\theta}{2+\theta})\\
\\
(\theta,1)
\end{array}\right.\right].
\end{eqnarray}
The contribution to the Barnes-Mellin integral from the residues vanish 
for large $|x|$ and the asymptotic form of the Fox function is 
stretched-exponential \cite{bra,sro2}: 
\begin{equation}
\label{expo}
p(x,t)\sim a^{1+\theta}|x|^\theta \exp[-c_2 (a|x|)^{2+\theta}], 
\end{equation}
where $a(t)$ is given by Eq.(\ref{a}) with $\mu=2$ and 
$c_2=(1+\theta)^{2+\theta}/(2+\theta)^{3+\theta}$. The variance can be easily 
evaluated:
\begin{equation}
\label{var}
\langle x^2\rangle=-\frac{\partial^2}{\partial k^2}{\widetilde p}(0,t)=
h_2a^{-2}, 
\end{equation}
where $h_2=\lim_{\mu\to2}h_\mu$. If $\lambda=0$, Eq.(\ref{var}) predicts the normal 
diffusion ($\theta=0$), subdiffusion ($\theta>0$), and superdiffusion 
($\theta<0$). Otherwise, the variance converges with time to a constant. 
Therefore, the parameter $\theta$ governs the transport speed. In the coupled 
CTRW, a large $\theta$ means that the average trapping time strongly rises 
with the distance. 

The system, which has been discussed above, 
is characterized by the same stability property as that 
of the driving noise: it is L\'evy distributed with the parameter $\mu$. 
Moreover, it reveals the simple scaling.  
One can ask whether the same properties hold for other 
systems driven by the L\'evy noise and a symmetric potential 
\cite{uwa}. The next case demonstrates that, even if 
the stability property is preserved, the index $\mu$ may change. 
Let us consider the drift in the form $F(x)=\lambda \hbox{sgn}(x)$, which corresponds 
to the wedge-shaped potential. We assume $0<\mu<1$, i.e. a process of the infinite 
mean, and $\mu+\theta>1$. The FPE is the following
\begin{equation}
\label{frace1}
\frac{\partial}{\partial t}p(x,t)=\lambda\frac{\partial}{\partial x}
[\hbox{sgn}(x)p(x,t)]+K^\mu\frac{\partial^\mu}{\partial |x|^\mu}[|x|^{-\theta}p(x,t)].
\end{equation}
Its cosine Fourier transform reads
\begin{equation}
\label{fracek1}
\frac{\partial}{\partial t}{\widetilde p}(k,t)=-\lambda k\frac{\partial}{\partial k}
{\cal F}_c[|x|^{-1}p(x,t)]-K^\mu |k|^\mu{\cal F}_c[|x|^{-\theta}p(x,t)] 
\end{equation}
and the factor $|x|^{-1}$, which results from the change of the sine to cosine 
transform, introduces a new scale. We take into account that double scaling 
by assuming the solution in the form $p(x,t)=N[p_1(x,t)+\alpha p_2(x,t)]$, where 
\begin{eqnarray} 
\label{s2}
p_i(x,t)=f_i(t) H_{2,2}^{1,1}\left[f_i(t) |x|\left|\begin{array}{c}
(a_1^{(i)},A_1^{(i)}),(a_2^{(i)},A_2^{(i)})\\
\\
(b_1^{(i)},B_1^{(i)}),(b_2^{(i)},B_2^{(i)})
\end{array}\right.\right],
\end{eqnarray}
and $\alpha$ is determined by the initial condition 
$p(x,0)=\delta(x)=[\delta(x)+\alpha\delta(x)]/(1+\alpha)$; we assume that 
$\alpha\ne0$ if $\lambda\ne0$. The solving method 
is similar to the previous case: we insert Eq.(\ref{s2}) into 
Eq.(\ref{fracek1}) and compare the terms. 
The essential point is to realize that the first term on rhs 
upgrades the index $\mu$ by one, because the expansions of 
${\cal F}_c[|x|^{-1}p_i(x,t)]$ are determined by the terms 
$|k|^{(2-a_1^{(i)})/A_1^{(i)}}$. Then 
${\cal F}_c[|x|^{-1}p_i(x,t)]\approx \hbox{const}-{h'}_\mu^{(1)}f_1^{-\mu}|k|^{\mu+1}$, 
where terms of the order $\mu+2$ have been neglected. 
We put $a_i^{(1)}=a_i$, $A_i^{(1)}=A_i$, $b_i^{(1)}=b_i$, and $B_i^{(1)}=B_i$. 
The coefficients for $p_2$ are the same except $\mu\to\mu+1$. 
By comparing the terms of order $\mu$ and $\mu+1$, we obtain a set 
of two differential equations 
\begin{eqnarray}
\label{eqa1}
\dot\xi_1&=&K^\mu[\frac{h_0^{(1)}}{h_\mu^{(1)}}\xi_1^{-\theta/\mu}+\alpha
\frac{h_0^{(2)}}{h_\mu^{(1)}}\xi_2^{-\theta/(\mu+1)}]\nonumber\\
&\nonumber\\
\dot\xi_2&=&(\mu+1)\frac{\lambda {h'}_\mu^{(1)}}{\alpha h_\mu^{(2)}}\xi_1,
\end{eqnarray}
where $\xi_1=f_1^{-\mu}$, $\xi_2=f_2^{-\mu-1}$, $h_\mu^{(1)}=h_\mu$, 
$h_\mu^{(2)}=h_{\mu+1}$, $h_0^{(1)}=h_0$, $h_0^{(2)}=h_0(\mu\to\mu+1)$, and 
${h'}_\mu^{(1)}=-N(\mu+\theta)\Gamma(-\mu-1)
\Gamma(2+\mu)\sin(\pi\mu/2)/\Gamma[1/2+(\mu+\mu\theta+2)/(2+\theta)]
\Gamma[-(\mu+\theta)/(2+\theta)]$. 
%; the normalization constant is 
%\begin{equation}
%\label{n1}
%N=\frac{1}{2}\frac{\Gamma[-\theta/(2+\theta)]\Gamma[1/2+2/
%(2+\theta)]}{\Gamma(1+\theta)\{\Gamma[-\theta/(\mu+\theta)]+
%\alpha\Gamma[-\theta/(\mu+\theta)]\}}.
%\end{equation}
\begin{figure}[tbp]
\includegraphics[width=8.5cm]{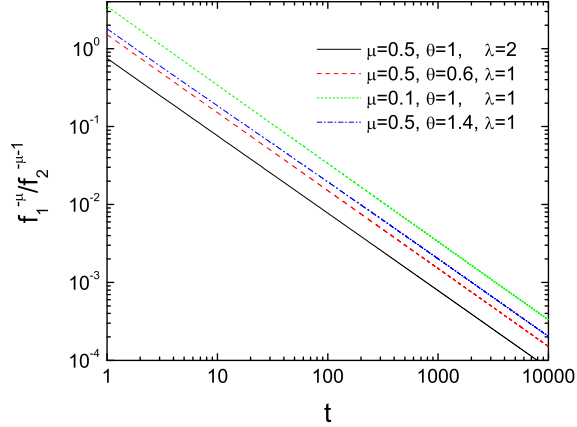}
\caption{(Colour online). The ratio which determines the relative 
contribution to the solution of Eq.(\ref{frace1}) from 
the terms which correspond to the L\'evy indexes $\mu$ and $\mu+1$, 
for various parameters of the process. All curves fall like $1/t$.}
\end{figure}

Let us assume $\lambda>0$. The asymptotic form of $p(x,t)$ 
involves contributions from both $\mu$ and 
$\mu+1$, $p(x,t)=c_1f_1^{-\mu}|x|^{-\mu-1}+c_2f_2^{-\mu-1}|x|^{-\mu-2}+
o(|x|^{-2\mu-\theta-1})$, where $c_1$ and $c_2$ are constants. Therefore, 
the long tails which correspond to index $\mu$ prevail at large 
distances and the mean value is infinite. However, in the limit of 
long time the relative contribution to $p(x,t)$ from $p_1$ 
diminishes for a given $x$ since $f_2(t)$ falls faster than $f_1(t)$. 
To demonstrate that, we need to estimate the ratio $\xi_1/\xi_2=
f_1^{-\mu}/f_2^{-\mu-1}$. That quantity is presented in Fig.1. Its time 
dependence can be very well reproduced by the function $1/t$ 
and this pattern is generic for all values of $\lambda$, $\mu$, 
and $\theta$. Consequently, the contribution from $p_1$, 
which originates from the 
L\'evy process with the order parameter $\mu$, gradually fades away. 
Instead, the term corresponding to $\mu+1$ dominates the distribution 
at large time and $p_1$ enters the asymptotic expression  
only with a small weight. 
\begin{figure}[tbp]
\includegraphics[width=8.5cm]{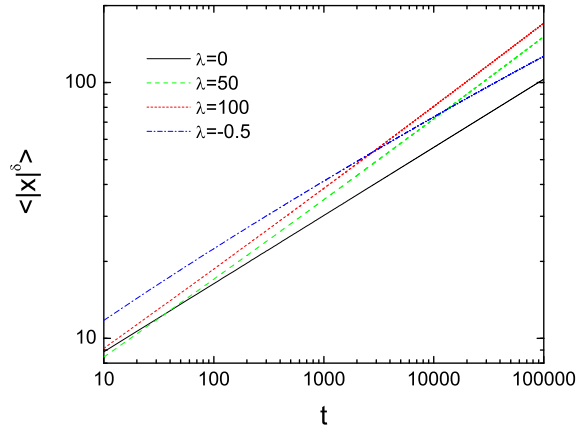}
\caption{(Colour online). The fractional moments calculated from 
Eq.(\ref{mom}) for $\delta=0.4$, 
$\mu=0.5$, $\theta=1$, and a few values of $\lambda$.}
\end{figure}

The transport in superdiffusive systems is used to be characterized by 
fractional moments of the order $\delta$, $\langle|x|^\delta\rangle$, where 
$\delta<\mu$, since all higher moments, in particular the variance, are 
divergent. That moment is easy to evaluate as the Mellin transform 
$\chi_i$ from the functions $p_i(x,t)$:
\begin{eqnarray}
\label{mom}
\langle|x|^\delta\rangle&=&2\int_0^\infty x^\delta p(x,t)dx=
2N[f_1^{-\delta}\chi_1(-\delta-1)+\alpha f_2^{-\delta}\chi_2(-\delta-1)]\nonumber\\
&\nonumber\\
&\sim&
%2N\frac{\Gamma(\theta+\delta+1)}
%{\Gamma(-\frac{\theta+\delta}{2+\theta})\Gamma(\frac{1}{2}+
%\frac{1-\theta+(1+\theta)(1+\delta)}{2+\theta})}
\Gamma(-\frac{\theta+\delta}{\mu+\theta})f_1^{-\delta}+
\alpha\Gamma(-\frac{\theta+\delta}{\mu+\theta+1})f_2^{-\delta}.
\end{eqnarray}
Fig.2 presents the fractional moments for a few values of the parameter $\lambda$. 
The dependence on time is algebraic and the power rises with $\lambda$; for 
$\lambda=0$ the moment is $\sim t^{\delta/(\mu+\theta)}$. 

Finally, let us mention the case of the attractive potential, $\lambda<0$. 
We require that also $\alpha<0$, in order to avoid shrinking of $p_2$ 
with time (c.f. second equation in Eq.(\ref{eqa1})), which is unphysical. 
The negative $\alpha$, in turn, results in negative $p(x,t)$ for large time, 
since $p_1$ falls faster with time than $p_2$. Therefore, the above solution 
of Eq.(\ref{frace1}) is correct for $\lambda<0$ only if time is not very large. 
The fractional moment rises slower than for $\lambda=0$ and also weaker 
than algebraically, which is presented in Fig.2. 

In summary, we have studied the stochastic systems in which 
the nonhomogeneous structure of the medium is reflected 
not only by an external potential but also directly by the random force 
in the form of the L\'evy distribution. 
Such systems are described by the fractional FPE with the variable 
diffusion coefficient; they have been solved in the limit of small wave numbers. 
The system can have the same stability properties 
as the driving noise; it is the case for the linear drift. 
However, the other drift we have considered, $\sim\hbox{sgn}(x)$, 
requires the additional L\'evy process and then the system has the double scaling. 
That second L\'evy distribution is characterized by larger order parameter 
and its weight rises with time.


\begin{thebibliography}{99}
\bibitem{uwa}
If $\theta=0$, the solution of FPE (\ref{fp}) for $F(x)=\hbox{const}$ is 
the L\'evy process with a translated coordinate \cite{jes}.
\end{thebibliography}
\end{document}